# Design of an all-facet illuminator for high NA EUV lithography exposure tool based on deep reinforcement learning

TONG LI,[1] YUQING CHEN,[1] YANQIU LI,[1,*] AND LIHUI LIU[1,*]

[1]*Key Laboratory of Photoelectronic Imaging Technology and System of Ministry of Education of China, School of Optics and Photonics, Beijing Institute of Technology, Beijing 100081, China*
*\* Corresponding author: liyanqiu@bit.edu.cn (Y. Li), liulihui@bit.edu.cn (L. Liu)*

**Abstract:** Using the illuminator for high numerical aperture (NA) extreme ultraviolet (EUV) exposure tool in EUV lithography can lead to support volume production of sub-2 nm logic nodes and leading-edge DRAM nodes. However, the typical design method of the illuminator has issues with the transmission owing to the limitation of optical structure that cannot further reduce process parameter $k_1$, and uniformity due to the restriction of matching method that can only consider one factor affecting uniformity. The all-facet illuminator can improve transmission by removing relay system. Deep reinforcement learning (RL) can improve the uniformity by considering multiple factors. In this paper, a design method of the all-facet illuminator for high NA EUV lithography exposure tool and a matching method based on deep RL for the double facets are proposed. The all-facet illuminator is designed using matrix optics, and removing relay system to achieve high transmission. The double facets is matched using the deep RL framework, which includes the policy network with improved trainability and low computational demands, and the reward function with great optimization direction and fast convergence rate, enabling to rapidly generate multiple matching results with high uniformity. An all-facet illuminator for a 0.55 NA EUV lithography exposure tool is designed by the proposed method. Simulation results indicate that the transmission is greater than 35%, and uniformity exceed 99% under multiple illumination pupil shapes.

## 1. Introduction

Extreme ultraviolet (EUV) lithography exposure tools are one of the key modules at semiconductor manufacturing. The latest generation of EUV lithography exposure tool increases resolution by increasing numerical aperture (NA) [1,2]. The underlying technology in this EUV lithography exposure tool is known as "High NA". The high NA EUV lithography exposure tool enables to support volume production of sub-2 nm logic nodes and leading-edge DRAM nodes [3]. To stick with the 6-inch mask, avoid the overlap of the light cones at the mask of incoming light and reflected light, and solve mask shadowing effect, the 8× demagnification in scan orientation is introduced while the 4× demagnification in the orientation perpendicular to scan orientation is kept. This is the so-called anamorphic design. A number of projection optics [4-14] and illuminators [15-19] for high NA EUV lithography exposure tool have been proposed in recent years, but few of these optics with a NA of 0.55 [10-13,15,19]. The 0.55 NA achieves an optimum balance among resolution, transmission, and field size, whereas the 0.50 NA only supports a quarter field [20,21] and the 0.60 NA has a low transmission [22]. Therefore, designing an illuminator for high NA, or 0.55 NA, EUV lithography exposure tool is crucial to achieve higher resolution.

The key objectives of the illuminator for high NA EUV lithography exposure tool are achieve high uniformity at the mask and switch different illumination pupil shapes. The typical illuminator is composed of two subsystems, the relay system and double facets. The double facets, which include the field facets and pupil facets that each consists of a large number of facet mirrors, enables to flexible switch the illumination pupil shape [23]. Relay system assists in establishing two optical conjugations in Köhler illumination: one is between the field facets and mask, and the other is between the pupil facets and exit pupil [24]. However, the typical illuminator limits the reduce in process parameter $k_1$ owing to its low transmission. In 2015, Mei *et al.* proposed a three-mirror illuminator to achieve higher transmission than the typical illuminator [25]. This illuminator consists of the field facets, the pupil facets, and a one-mirror relay system. However, the above illuminator is limited due to the surface shape of all pupil facet mirrors are fitted into freeform surface to improve the uniformity, which increases manufacturing cost. In 2025, Chen *et al.* proposed a glancing-incidence illuminator to improve the transmission [19]. However, this illuminator is limited because the size of the glancing-incidence mirror is too large to fabricate. Therefore, it is of great significance to design an illuminator for high NA EUV lithography exposure tool with high transmission, in order to reduce $k_1$ and thereby increase resolution.

The illumination pupil shapes are switched by first determining the distribution of the pupil facet mirrors that contribute to the illumination, and then adjusting the tilt angles of all field facet mirrors and the pupil facet mirrors that contribute to the illumination to transform the assignment relations between them [23]. The typical matching method matches each field facet mirror to one pupil facet mirror that is unmatched in a geometrically similar region [26]. However, this matching method is empirically determined, which makes it subjective and inefficient. The algorithm-based matching method can be considered as a solution to the above issues, as it can establish quantitative objective metrics and effectively generate matching results. In 2018, Jiang *et al.* proposed a matching method based on Kuhn-Munkres algorithm [16]. However, the matching method is limited by considering only one factor affecting uniformity, which ignores other factors. In 2025, Chen *et al.* proposed a matching method based on the multiparameter automatic alignment assignment model [19]. This matching method established a cost matrix that its elements were calculated from three factors affecting uniformity, and used Kuhn-Munkres algorithm to obtain a matching result. However, the above matching method can still generate only one matching result, which may not be the optimal uniformity. Hence, it is essential to design a matching method that can consider multiple factors affecting uniformity and generate multiple matching results with more optimal high uniformity.

In this paper, a design method of the all-facet illuminator for high NA EUV lithography exposure tool and a matching method based on deep RL for the double facets are proposed. The all-facet illuminator is calculated its coaxial initial structure first based on the two optical conjugations that achieves Köhler illumination using matrix optics, and its off-axis structure next by tilting and decentering. The all-facet illuminator can achieve high transmission by removing relay system. Deep RL can assign multiple factors into reward function and learn to achieve more optimal results by maximizing rewards. The use of deep RL in optics is considered to be a revolution as several fields such as imaging optical design [27], image reconstruction [28], and laser manufacturing [29]. The deep RL framework for the double facets matching includes the policy network that consists of a stochastic module and a deep learning module with improved trainability and low computational demands, and the reward function that assigns to three factors affecting uniformity with great optimization direction and fast convergence rate. The matching method based on deep RL can consider multiple factors affecting uniformity and generate multiple matching results with high uniformity. An all-facet illuminator for a 0.55 NA EUV lithography exposure tool is designed by the propose method. The simulation results of this illuminator demonstrate that the transmission is 35.32%, which is at least 39% higher than other illuminators for EUV lithography exposure tool, and uniformity can reach above 99% under multiple illumination pupil shapes.

## 2. Design method of all-facet illuminator

The structure of an illuminator is a critical factor to achieve high transmission. Since the exposure slit at the mask and the exit pupil are obtained from a projection optics for high NA EUV lithography exposure tool, the illuminator can be designed in the reverse sequence, in which the mask serves as the object plane and the exit pupil serves as the aperture stop. To reduce $k_1$ and thereby increase resolution, a design method of the all-facet illuminator for high NA EUV lithography exposure tool is proposed.

### 2.1. Basic of matrix optics

Matrix optics is a method that describes optical systems in the paraxial approximation. This method divides optical systems into two linearized relations: refraction of a ray and propagation of a ray, which can be written respectively in matrix form as:

$$\begin{bmatrix} y' \\ u' \end{bmatrix} = \begin{bmatrix} 1 & 0 \\ \frac{n-n'}{n'r} & \frac{n}{n'} \end{bmatrix} \begin{bmatrix} y \\ u \end{bmatrix}, \tag{1}$$

$$\begin{bmatrix} y' \\ u' \end{bmatrix} = \begin{bmatrix} 1 & d \\ 0 & 1 \end{bmatrix} \begin{bmatrix} y \\ u \end{bmatrix}, \tag{2}$$

where $y$ and $y'$ are the paraxial ray heights, $n$ and $n'$ are the refractive indices, $u$ and $u'$ are the paraxial ray angles, and $d$ is the traveled distance. By Eq. (1) and Eq. (2), the refraction matrix $R$ and the translation matrix $T$ are respectively defined as:

$$R = \begin{bmatrix} 1 & 0 \\ \frac{n-n'}{n'r} & \frac{n}{n'} \end{bmatrix}, \tag{3}$$

$$T = \begin{bmatrix} 1 & d \\ 0 & 1 \end{bmatrix}. \tag{4}$$

From Eq. (1) to Eq. (4), the linearized relation that a ray pass through an optical system with $k$ surfaces can be written in matrix form as:

$$\begin{bmatrix} y'_k \\ u'_k \end{bmatrix} = R_k T_{k-1} R_{k-1} \cdots T_1 R_1 \begin{bmatrix} y_1 \\ u_1 \end{bmatrix}, \tag{5}$$

By Eq. (5), the matrix $S$, which describes the optical arrangement formed by an optical system, is defined as:

$$S = R_k T_{k-1} R_{k-1} \cdots T_1 R_1 = \begin{bmatrix} A & B \\ C & D \end{bmatrix}, \tag{6}$$

where $A$, $B$, $C$, and $D$ are the Gaussian constants. By the properties of matrix, the determinant of the product of multiple matrices is equal to the product of the determinants of multiple matrices. Therefore, the Gaussian constants in the matrix $S$ satisfies the following equation:

$$AD - BC = 1. \tag{7}$$

From Eq. (6), the matrix $M$, which describes the linearized relation for a ray pass through a pair of conjugate planes, is defined as:

$$M = \begin{bmatrix} 1 & d_i \\ 0 & 1 \end{bmatrix} \begin{bmatrix} A & B \\ C & D \end{bmatrix} \begin{bmatrix} 1 & d_o \\ 0 & 1 \end{bmatrix} = \begin{bmatrix} \beta & 0 \\ \alpha & 1/\beta \end{bmatrix}, \quad (8)$$

where $d_o$ is the object distance, $d_i$ is the image distance, $\beta$ is the lateral magnification, and $\alpha$ is the longitudinal magnification.

### 2.2. Initial system calculation

The all-facet illuminator only consists of double facets, which approximately is a two-mirror structure. In comparison to the typical illuminator with a four-mirror structure, the all-facet illuminator removes relay system to improve the transmission. However, the optical conjugations in Köhler illumination are disabled. To solve this issue, the pupil facets is coincident with the exit pupil to establish the optical conjugation between the pupil facets and exit pupil, and using the pupil facets to establish the optical conjugation between the field facets and mask.

In the all-facet illuminator, all pupil facet mirrors are arranged in the exit pupil. To determine the coordinate and size of each pupil facet mirror, the exit pupil is pixelated. The original and pixelated exit pupil is shown in Fig. 1. The coordinate origin of each pupil facet mirror is at the center of each grid in the pixelated exit pupil, and its size is limited within the grid.

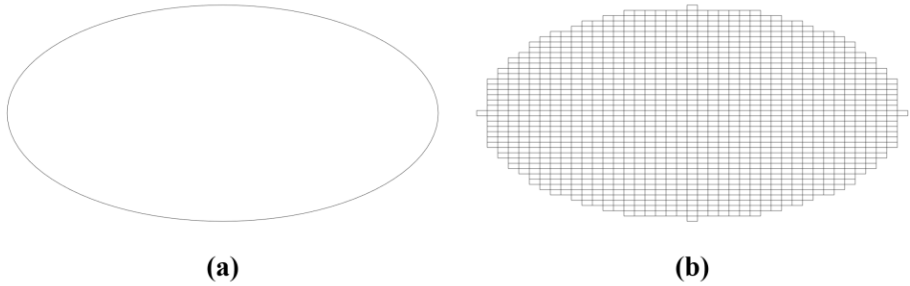

**Fig. 1.** (a) The original exit pupil. (b) The pixelated exit pupil.

The radius of the pupil facet mirror and field facet mirror are determined in the reverse sequence using matrix optics. The coaxial structure of part all-facet illuminator from the mask to the field facets is depicted in Fig. 2, in which $l_m$ is the distance from the pupil facets to the mask, $l'_m$ is the distance from the pupil facets to the field facets, and $r_{PF}$ is the radius of the pupil facet mirror. Owing to EUV is absorbed strongly by all substances, EUV lithography must take place in vacuum environment and the illuminator for EUV lithography exposure tool must adopt the reflective structure [30]. Therefore, the index of refraction $n = 1$ and index of refraction $n' = -1$. The mask and field facets are conjugate, thus the matrix $S$ of this coaxial structure is:

$$S_1 = R_{PF} = \begin{bmatrix} 1 & 0 \\ -2/r_{PF} & -1 \end{bmatrix} = \begin{bmatrix} A_1 & B_1 \\ C_1 & D_1 \end{bmatrix}, \quad (9)$$

and the following equation is obtained:

$$C_1 = -2/r_{PF}. \quad (10)$$

By Eq. (9), the matrix $M$ of this coaxial structure is:

$$M_1 = T'_m S_1 T_m$$
$$= \begin{bmatrix} 1 & -l'_m \\ 0 & 1 \end{bmatrix} \begin{bmatrix} A_1 & B_1 \\ C_1 & D_1 \end{bmatrix} \begin{bmatrix} 1 & -l_m \\ 0 & 1 \end{bmatrix}$$
$$= \begin{bmatrix} A_1 - l'_m C_1 & -l_m A_1 + B_1 + l_m l'_m C_1 - l'_m D_1 \\ C_1 & -l_m C_1 + D_1 \end{bmatrix}$$
$$= \begin{bmatrix} \beta_m & 0 \\ \alpha_m & 1/\beta_m \end{bmatrix}, \tag{11}$$

where $\beta_m$ is the lateral magnification of the mask, and $\alpha_m$ is the longitudinal magnification of the mask. The equation is derived as follows:

$$-l_m C_1 + D_1 = 1/\beta_m. \tag{12}$$

By converting and combining Eq. (10) and Eq. (12), $r_{PF}$ can be calculated as:

$$r_{PF} = \frac{2\beta_m l_m}{1-\beta_m}. \tag{13}$$

Among the parameters in this coaxial structure, $\beta_m$ is determined by the position of the field facets, while its position is calculated using ray tracing from its size, and $l_m$ has been obtained from the projection optics.

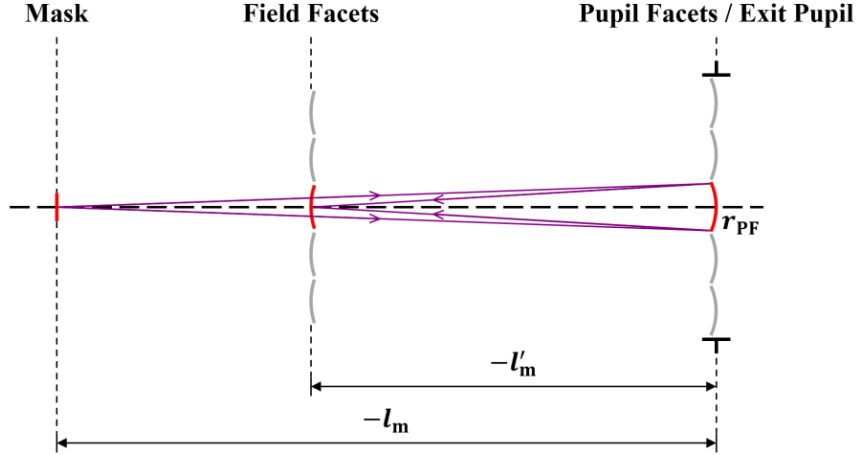

Fig. 2. The coaxial structure of part all-facet illuminator from the mask to the field facets.

The coaxial structure of part all-facet illuminator from the pupil facets to the intermediate focus is shown in Fig. 3, in which $l_{PF}$ is the distance from the field facets to the pupil facets, $l'_{PF}$ is the distance from the field facets to the intermediate focus, and $r_{FF}$ is the radius of the field facet mirror. The pupil facets and intermediate are conjugate, thus the matrix $S$ of this coaxial structure is:

$$S_2 = R_{FF} = \begin{bmatrix} 1 & 0 \\ -2/r_{FF} & 1 \end{bmatrix} = \begin{bmatrix} A_2 & B_2 \\ C_2 & D_2 \end{bmatrix}, \tag{14}$$

and the following equation is obtained:

$$C_2 = -2/r_{FF}. \tag{15}$$

By Eq. (14), the matrix $M$ of this coaxial structure is:

$$M_2 = T'_{PF} S_2 T_{PF}$$

$$= \begin{bmatrix} 1 & l'_{PF} \\ 0 & 1 \end{bmatrix} \begin{bmatrix} A_2 & B_2 \\ C_2 & D_2 \end{bmatrix} \begin{bmatrix} 1 & l_{PF} \\ 0 & 1 \end{bmatrix}$$

$$= \begin{bmatrix} A_2 + l'_{PF} C_2 & l_{PF} A_2 + B_2 + l_{PF} l'_{PF} C_2 + l'_{PF} D_2 \\ C_2 & l_{PF} C_2 + D_2 \end{bmatrix}$$

$$= \begin{bmatrix} \beta_{PF} & 0 \\ \alpha_{PF} & 1/\beta_{PF} \end{bmatrix}, \quad (16)$$

where $\beta_{PF}$ is the lateral magnification of the pupil facets, and $\alpha_{PF}$ is the longitudinal magnification of the pupil facets. The equation is derived as follows:

$$l_{PF} C_2 + D_2 = 1/\beta_{PF}. \quad (17)$$

By converting and combining Eq. (15) and Eq. (17), $r_{FF}$ can be calculated as:

$$r_{FF} = \frac{2\beta_{PF} l_{PF}}{\beta_{PF} - 1}. \quad (18)$$

Within the above parameters in this coaxial structure, $\beta_{PF}$ is determined once $r_{PF}$ has been calculated, and $l_{PF}$ is equal to $-l'_m$, which has been obtained from the projection optics.

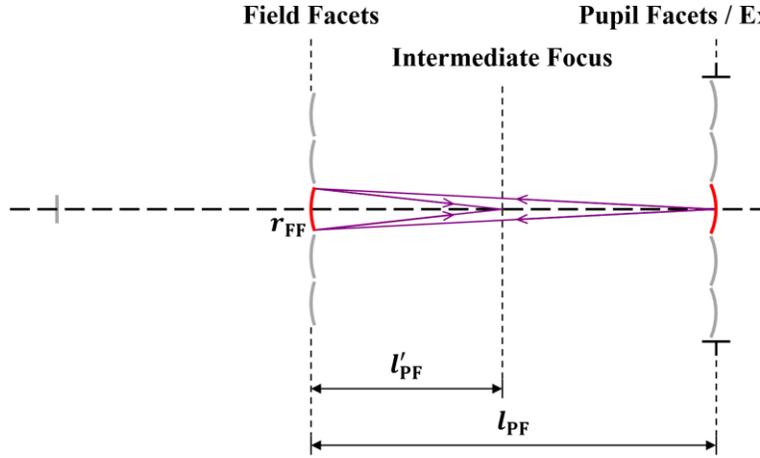

**Fig. 3.** The coaxial structure of part all-facet illuminator from the pupil facets to the intermediate focus.

After the radius of the pupil facet mirror and field facet mirror are obtained, the initial system with coaxial structure of the all-facet illuminator that determines the position of each component can be calculated.

### 2.3. Off-axis structure establishment

To eliminate ray obscuration, the coaxial initial structure of the all-facet illuminator must be tilted and decentered. In particular, the pupil facets and the exit pupil are only permitted to decenter along the $y$ axis and are forbidden to tilt as these components are determined from the entrance pupil of the projection optics, which is perpendicular to the optical axis. The off-axis structure of the all-facet illuminator in the reverse sequence is shown in Fig. 4.

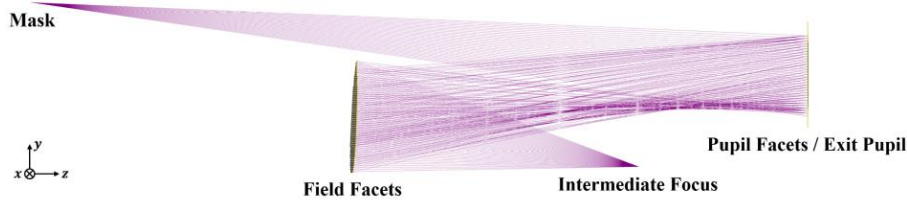

**Fig. 4.** The off-axis structure of the all-facet illuminator.

The input and output vectors of each facet mirror in the double facets are determined using ray tracing. As illustrated in Fig. 5, the input vector and output vector of the $j_{th}$ pupil facet mirror $I_{PF_j}$ and $O_{PF_j}$, and the input vector and output vector of the $i_{th}$ field facet mirror $I_{FF_i}$ and $O_{FF_i}$ is obtained by tracing the ray from the central field point to the intermediate focus. Afterwards, calculating the normal unit vector of the $j_{th}$ pupil facet mirror $N_{PF_j}$ and $i_{th}$ field facet mirror $N_{FF_i}$, respectively, through these vectors by the following equation:

$$N = \frac{O - I}{\|O - I\|_2}, \tag{19}$$

where $\|\cdot\|_2$ is the L2 norm, or the Euclidean norm, which measures the magnitude of a vector.

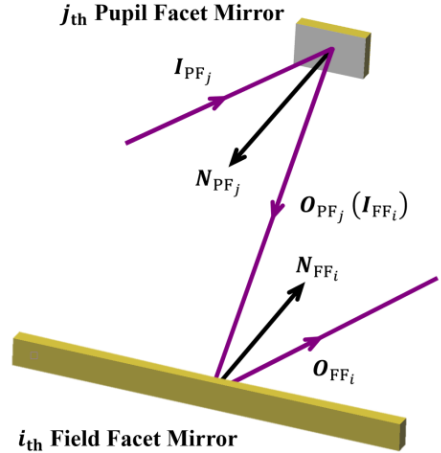

**Fig. 5.** The schematic diagram of calculating the normal unit vector of the $j_{th}$ pupil facet mirror and $i_{th}$ field facet mirror.

Once $N_{PF_j}$ and $N_{FF_i}$ are obtained, the tilt angles of the $j_{th}$ pupil facet mirror and $i_{th}$ field facet mirror can be calculated by Euler's rotation theorem. The field and pupil facet mirrors can rotate about the $x$ and $y$ axes, the rotation matrix $R$ is given as:

$$R = R_x(\alpha) R_y(\beta) = \begin{bmatrix} \cos\beta & 0 & \sin\beta \\ \sin\alpha\sin\beta & \cos\alpha & -\sin\alpha\cos\beta \\ -\cos\alpha\sin\beta & \sin\alpha & \cos\alpha\cos\beta \end{bmatrix}, \tag{20}$$

where $\alpha$ is the Euler angle about the $x$ axis, and $\beta$ is the Euler angle about the $y$ axis. As the third column of the matrix $R$ represents the unit vector of the rotated $z$ axis in the original coordinate system, the tilt angles of the $j_{th}$ pupil facet mirror and $i_{th}$ field facet mirror are calculated through the elements of $\mathbf{N}_{PF_j}$ and $\mathbf{N}_{FF_i}$ using inverse trigonometric functions.

## 3. Matching method of double facets

The double facets is a critical component of the illuminator for high NA EUV lithography exposure tool to achieve multiple illumination pupil shapes. The assignment relations between all field facet mirrors and the pupil facet mirrors that contribute to the illumination are determined based on a certain illumination pupil shape. To consider multiple factors affecting uniformity and generate multiple matching results with high uniformity, a matching method based on deep RL is proposed.

### 3.1. Overall framework for double facets matching

The deep RL framework for the double facets matching is shown in Fig. 6. This deep RL framework uses Double Deep Q-Network (DDQN), which is a deep RL algorithm that combines double Q-Learning with deep neural network, enabling it to output a vector of action values with reduced overestimations for a given state [31]. For the double facets matching, the environment is all field facet mirrors and the pupil facet mirrors that contribute to the illumination, the agent is the system that determines the matching result, the state $s$ is a matching status of all field facet mirrors and the pupil facet mirrors that contribute to the illumination, the action $a$ is a pair of index numbers corresponding to one field facet mirror and one pupil facet mirror that are selected according to the state $s$. The deep RL framework for the double facets matching is based on the standard RL process, which consists of the following steps: The agent selects an action $a$ for a state $s$ from the environment, then inputs this action $a$ to the environment. The environment interprets the action $a$ into a reward $r$ and a next state $s'$, then feeds this reward $r$ and this next state $s'$ back to the agent.

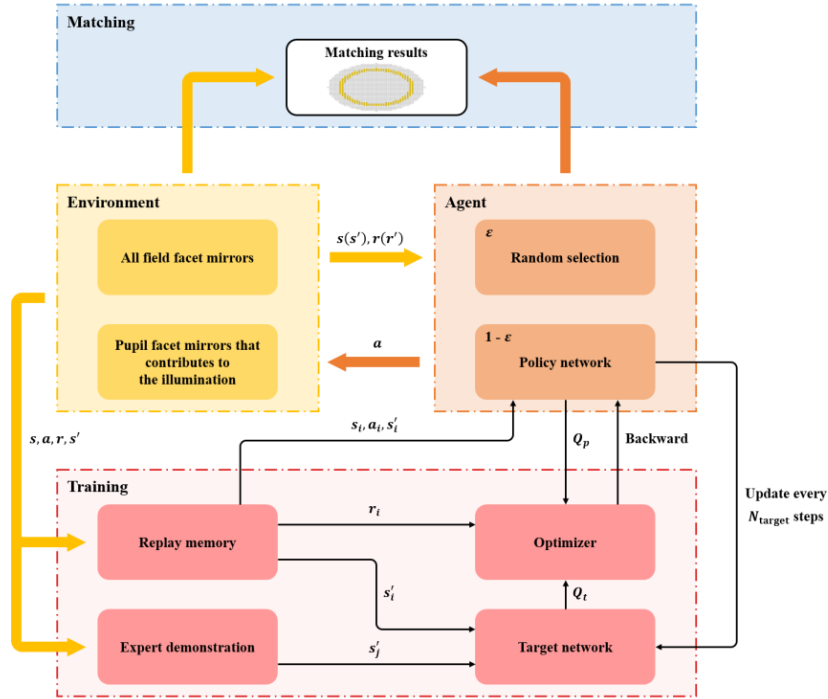

**Fig. 6.** The deep RL framework for the double facets matching

The agent selects actions based on $\varepsilon$-greedy strategy, in which actions are selected through the random selection with probability $\varepsilon$, or through the policy network with probability $1-\varepsilon$. The policy network can approximate the action value $Q(s,a)$, which represents the expected cumulative reward by selecting the action $a$ in the state $s$. The policy network is trained with the assistance of target network and replay memory. The target network is a separate copy of the policy network that is updated with a fixed frequency, which generates more stable Q-value targets for training and thereby reduces training volatility. The replay memory stores the transitions observed by the agent, enabling the policy network to randomly sample these transitions from it for updating, which greatly stabilizes and improves the training procedure. A single transition $(s_i, a_i, r_i, s_i')$ consists of the state $s$, the action $a$, the reward $r$, and the next state $s'$. In training stage, a batch of transitions are randomly sampled from the replay memory and calculate the target Q-value for each experience by the Bellman Equation:

$$a'_{max} = \mathrm{argmax}_{a'} Q_e(s_i', a'), \tag{21}$$

$$Q_i = \begin{cases} r_i, & \text{if the episode ends at } s_i' \\ r_i + \gamma Q_t(s_i', a'_{max}), & \text{otherwise} \end{cases}, \tag{22}$$

where $Q_e(s_i', a')$ and $Q_t(s_i', a'_{max})$ are the Q-value predicted by the policy network and the target network, respectively, $Q_i$ is the target Q-value, $\gamma$ is the discount rate. The loss $L$ between $Q_i$ and $Q_e(s_i', a')$ is calculated, and gradient descent is applied with this loss to update the weights of the policy network until the loss $L$ converges or no longer decreases significantly. Once the policy network training is completed and achieving high performance, the agent can generate multiple matching results with high uniformity under multiple illumination pupil shapes in matching stage.

*3.2. Policy network architecture*

The policy network in an agent is a neural network that approximates the optimal action-value function and maps states to actions. The architecture of the policy network is crucial, as it determines the trainability and computational demands.

The architecture of the policy network in the deep RL framework for the double facets matching is shown in Fig. 7. The policy network consists of one stochastic module and one deep learning module. The stochastic module performs a random selection process. The deep learning module employs a multilayer perceptron (MLP), which is composed of multiple layers of neurons. The MLP consists of an input layer, multiple hidden layers, and an output layer. Each hidden layer receives input from the previous layer, and the output layer receives input from the last hidden layer. The number of neurons in each layer is determined by the number of all field facet mirrors and the pupil facet mirrors that contribute to the illumination, demonstrating that it depends on the number of the double facets. Each module accepts a vector as input, and produces a scalar as output. For the stochastic module, the input vector represents the matching status of all field facet mirrors, and the output scalar represents one field facet mirror selected in current state. For the deep learning module, the input vector is a concatenation of two vectors. The first vector represents the selected one field facet mirror using one-hot encoding, and the second vector represents the matching status of the pupil facet mirrors that contribute to the illumination. The output scalar represents one pupil facet mirror matched with the one field facet mirror selected in current state. The element of each input vector is assigned a value of 1 or 0, where 1 represents that the corresponding field facet mirror or pupil facet mirror is available for matching and 0 represents that it has already been matched. The policy network accepts a matching status of all field facet mirrors and the pupil facet mirrors that contribute to the illumination, and produces a pair of index numbers corresponding to one field facet mirror and one pupil facet mirror that are selected according to the matching status.

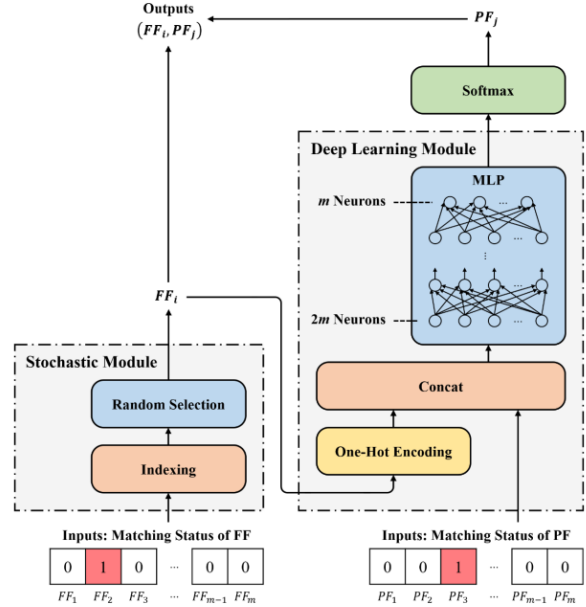

**Fig. 7.** The architecture of the policy network in the deep RL framework. The red-colored element represents that the status of the corresponding field facet or pupil facet is matched in previous step.

Employing the above architecture of the policy network in the deep RL framework for the double facets matching improves trainability and lowers computational demands. The agent can determine more optimal assignment relations between all field facet mirrors and the pupil facet mirrors that contribute to the illumination.

### 3.3. Reward function setting

The reward function is a quantitative criterion that dictates how an agent learns and behaves by evaluating the quality of actions taken in states. The reward function is essential, as it directly influences the optimization direction and convergence rate of deep RL.

To consider more factors affecting uniformity, the reward function in the deep RL framework for the double facets matching is assigned to multiple factors affecting uniformity. An illuminator for EUV lithography exposure tool with the simplified double facets is shown in Fig. 8. This illuminator consists of five pupil facet mirrors located at the +x edge, -x edge, +y edge, -y edge, and center, respectively, and one field facet mirror at the center. The illumination spots at the mask obtained by five individual channels are shown in Fig. 9. The illumination spot obtained by the channel that includes pupil facet mirrors located at the +x edge or -x edge exhibit significant tilt and slight decenter, those at the +y edge or -y edge exhibit large decenter, while the one at the center exhibit no tilt or decenter. The results indicate that tilt and decenter exist in the illumination spot and manifest at varying numerical values. These factors can severely decrease the uniformity, as the illumination spot of each channel are superimposed on one another at the mask to achieve high uniformity. Therefore, the tilt of the illumination spot, the decenter of the illumination spot in the $x$ direction, and the decenter of the illumination spot in the $y$ direction are three factors affecting uniformity and are assigned into the reward function in the deep RL framework for the double facets matching.

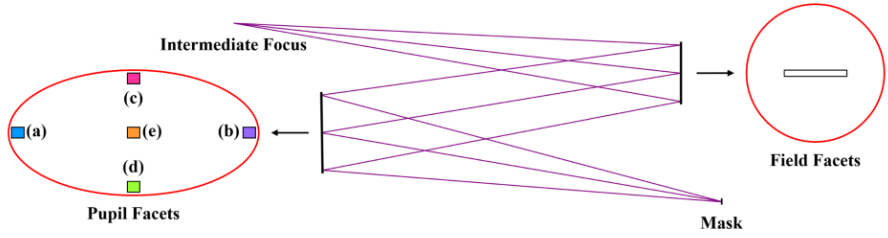

**Fig. 8.** The illuminator for EUV lithography exposure tool with the simplified double facets. (a) The -x edge pupil facet. (b) The +x edge pupil facet. (c) The +y edge pupil facet. (d) The -y edge pupil facet. (e) The center pupil facet.

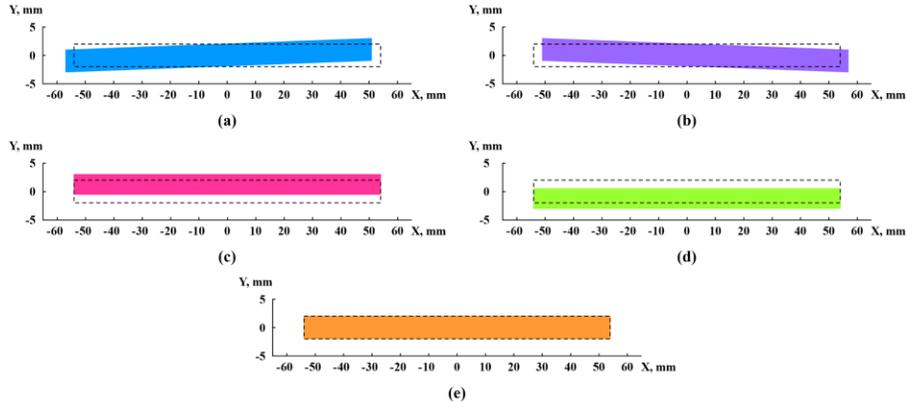

**Fig. 9.** The illumination spots at the mask by five individual channels that each consists of the center field facet mirror and one pupil facet mirror in Fig. 8. The black dotted rectangle is the ideal light spot at the mask.

The numerical ranges of the three factors affecting uniformity are shown in Fig. 10 (a). Since the magnitude of numerical values of these factors differ significantly, directly using these factors in the reward function may cause a certain factor to dominate the reward. To address this issue, min-max normalization is applied to unify the numerical range of these factors. Min-max normalization rescales the original values to a specified target range. The formula for min-max normalization within the range $[-1,1]$ is expressed as follows:

$$v'_i = -1 + \frac{2(v_i - \min(V))}{\max(V) - \min(V)}, \qquad (23)$$

where $V$ is the set of original values, $v_i$ is the original value, and $v'_i$ is the normalized value. The normalized numerical ranges of these factors are shown in Fig. 10 (b). These factors can be generally considered to exert equal influence on the reward after normalization.

The reward function in the deep RL framework for the double facets matching has three forms: if the matching is incomplete, the reward function is formulated as:

$$reward = 0, \qquad (24)$$

if the constraint that enables only one assignment relation between each field facet mirror and each pupil facet mirror that contributes to the illumination is not satisfied, the matching is terminated immediately. In this situation, the reward function is defined as:

$$reward = -10, \qquad (25)$$

if the matching is complete, the reward function is formulated as:

$$reward = -\text{Avg}\left(\sum_{i=1}^{M}\sum_{j=1}^{M}|TILT_{i,j}|\right) - \text{Avg}\left(\sum_{i=1}^{M}\sum_{j=1}^{M}|XDE_{i,j}|\right) - \text{Avg}\left(\sum_{i=1}^{M}\sum_{j=1}^{M}|YDE_{i,j}|\right). \quad (26)$$

where $TILT_{i,j}$ is the tilt of the illumination spot obtained by the channel that consists of the $i_{th}$ field facet mirror and the $j_{th}$ pupil facet mirror, $XDE_{i,j}$ is the decenter of the illumination spot in the $x$ direction, $YDE_{i,j}$ is the decenter of the illumination spot in the $y$ direction, and $\text{Avg}(\cdot)$ is the function that calculates the average of numerical values.

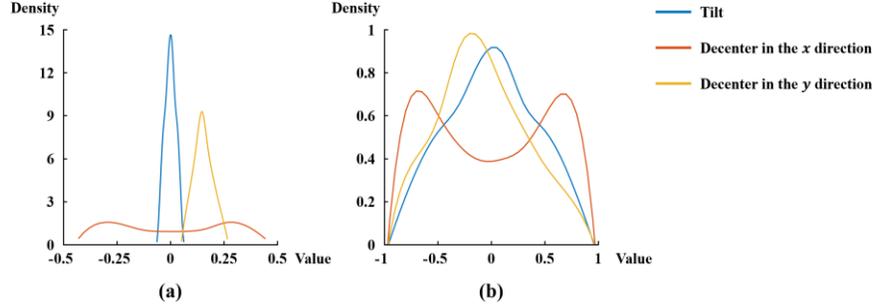

**Fig. 10.** (a) The numerical ranges of the three factors affecting uniformity before normalizing. (b) The numerical ranges of the three factors affecting uniformity after normalizing.

Using the above reward function in the deep RL framework for the double facets matching leads to better optimization direction and faster convergence rate. The agent can consider more factors affecting uniformity and generate matching results with high uniformity.

### 3.4. Matching process

The matching process of the illuminator using the matching method based on deep RL is shown in Fig. 11. The index numbers of all field facet mirrors and the pupil facet mirrors that contribute to the illumination, as well as the tilt of the illumination spots and the decenter of the illumination spots obtained by channels that each consists of one field facet mirror and one pupil facet mirror that contributes to the illumination without duplication, can be calculated based on the illuminator and a certain illumination pupil shape. The agent generates multiple matching results with high uniformity that consider the above parameters. These matching results can be filtered and sorted by designers on the basis of their own needs. Compared with existing matching methods, the matching method based on deep RL can consider multiple factors affecting uniformity and generate multiple matching results with more optimal high uniformity.

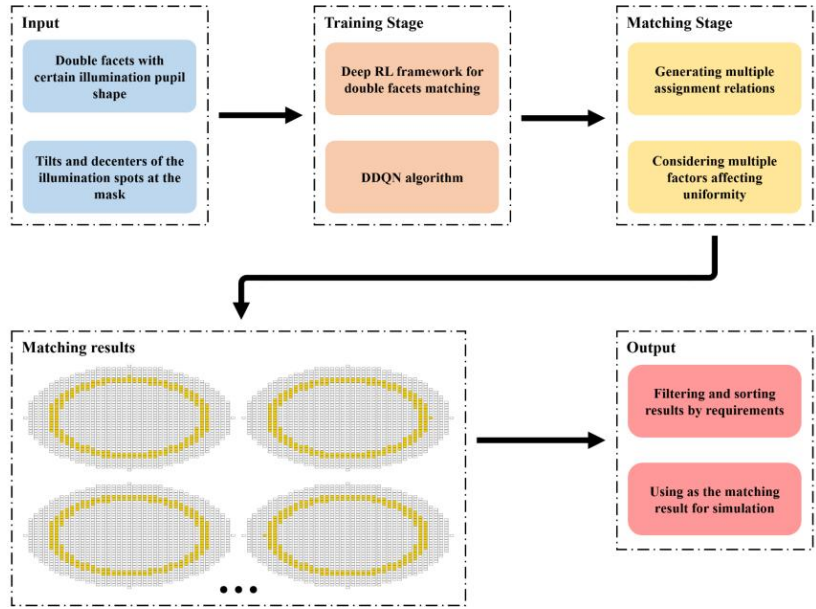

**Fig. 11.** The matching process of the illuminator using the matching method based on deep RL.

## 4. Results

An all-facet illuminator for a 0.55 NA EUV lithography exposure tool is designed to verify the effect of the proposed methods. The layout of this illuminator is shown in Fig. 12, and its parameters are listed in Table 1.

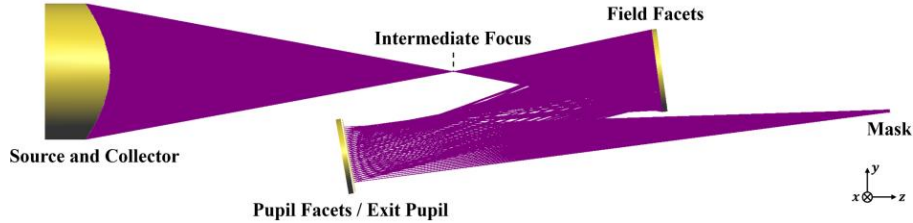

**Fig. 12.** The layout of the all-facet illuminator for a 0.55 NA EUV lithography exposure tool.

**Table 1** The parameters of the all-facet illuminator for a 0.55 NA EUV lithography exposure tool.

| Parameter | Specification |
| --- | --- |
| Numerical aperture | 0.55 |
| Wavelength (nm) | 13.5 |
| Demagnification | 4 (X) / 8 (Y) |
| Exposure slit size (mm) | 104 × 4 |
| Exit pupil diameter (mm) | 573.4608 (X) × 286.7304 (Y) |
| Exit pupil distance (mm) | 2154.7019 |

The placement of the field facets and pupil facets is shown in Fig. 13. All field facet mirrors are arranged in an ellipse to avoid beam obscuration, while all pupil facet mirrors are arranged in the elliptical exit pupil to achieve the optical conjugation between the pupil facets and exit pupil. There are 214 field facet mirrors and 1,249 pupil facet mirrors in total, each field facet mirror with dimension of 70 mm × 5 mm and each pupil facet mirror with dimension of 10 mm

× 6 mm. The surface shapes of all field facet mirrors and all pupil facet mirrors are spherical, each with the radius that can be easily calculated and fabricated.

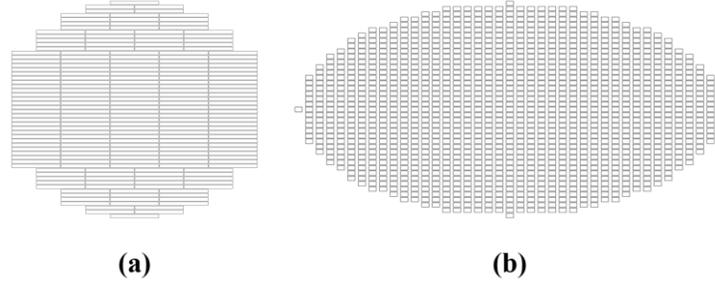

**Fig. 13.** (a) The placement of the field facets. (b) The placement of the pupil facets.

The MLP in the deep RL framework for the double facets matching consists of four layers of neurons: an input layer with 384 neurons, a first hidden layer with 512 neurons, a second hidden layer with 256 neurons, and an output layer with 192 neurons. The optimizer, activation function, loss function, capacity of the replay memory, probability of random selection $=\varepsilon$, and discount rate $\gamma$ were Adam, ReLU, MSE, 500, 0.05, and 0.9, respectively. The training was performed on one machine with one NVIDIA GeForce RTX 3090 24 GB GPU. Multiple matching results under a certain illumination pupil shape generated by the matching method based on deep RL are shown in Fig. 14, where $Avg.TILT$ is the average tilt of the illumination spot, $Avg.XDE$ is the average decenter of the illumination spot in the $x$ direction, and $Avg.YDE$ is the average decenter of the illumination spot in the $y$ direction.

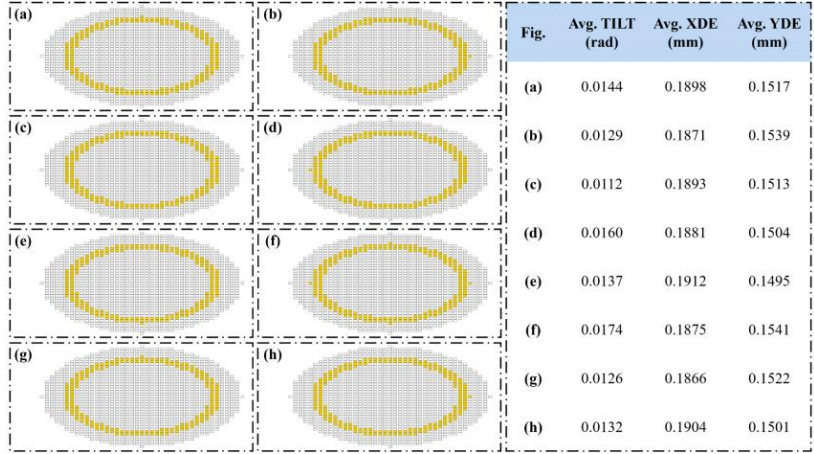

| Fig. | Avg. TILT (rad) | Avg. XDE (mm) | Avg. YDE (mm) |
|---|---|---|---|
| (a) | 0.0144 | 0.1898 | 0.1517 |
| (b) | 0.0129 | 0.1871 | 0.1539 |
| (c) | 0.0112 | 0.1893 | 0.1513 |
| (d) | 0.0160 | 0.1881 | 0.1504 |
| (e) | 0.0137 | 0.1912 | 0.1495 |
| (f) | 0.0174 | 0.1875 | 0.1541 |
| (g) | 0.0126 | 0.1866 | 0.1522 |
| (h) | 0.0132 | 0.1904 | 0.1501 |

**Fig. 14.** Multiple matching results under a certain illumination pupil shape generated by the matching method based on deep RL. Using annular illumination pupil shape as example.

The distribution of the pupil facet mirrors that contribute to the illumination under six illumination pupil shapes, which are annular, dipole, dipole 90°, quasar, quasar 45°, and leaf, is shown in Fig. 15, where the yellow-colored pupil facet mirrors are the pupil facet mirrors that contribute to the illumination will in the illumination. The number of pupil facet mirrors that contribute to the illumination under all illumination pupil shapes always equals that of all field facet mirrors. The parameters of the six illumination pupil shapes are listed in Table 2, in which $\sigma_{in}$ is the inner partial coherence factor, $\sigma_{out}$ is the outer partial coherence factor, and $\theta$ is the opening angle.

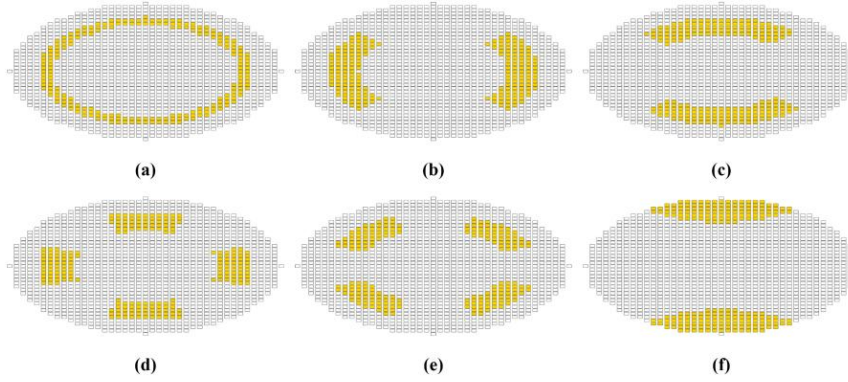

**Fig. 15.** The distribution of the pupil facet mirrors that contribute to the illumination under six illumination pupil shapes. (a) Annular illumination pupil shape. (b) Dipole illumination pupil shape. (c) Dipole 90° illumination pupil shape. (d) Quasar illumination pupil shape. (e) Quasar 45° illumination pupil shape. (f) Leaf illumination pupil shape.

**Table 2** The parameters of the six illumination pupil shapes.

| Pupil Shape | $\sigma_{in}$ | $\sigma_{out}$ | $\theta$ |
|---|---|---|---|
| Annular | 0.68 | 0.80 | - |
| Dipole | 0.54 | 0.80 | 0° |
| Dipole 90° | 0.54 | 0.80 | 90° |
| Quasar | 0.52 | 0.80 | 0° |
| Quasar 45° | 0.52 | 0.80 | 45° |
| Leaf | - | - | - |

The total luminous flux delivered to the illumination plane is one of the primary concerns for illuminator. The transmission of the all-facet illuminator for high NA EUV lithography exposure tool is defined as:

$$Transmission = \eta_{collecting} \times \eta_{FF} \times \eta_{PF} \times \eta_{mask}, \tag{27}$$

where $\eta_{collecting}$ is the efficiency of the field facets to collect the incident beam, $\eta_{FF}$ is the transmission of the field facets, $\eta_{PF}$ is the transmission of the pupil facets, and $\eta_{mask}$ is the ratio of the light irradiance within the exposure slit to the light irradiance at the mask. The transmission of the multilayer-coated surface for normal-incidence light is set to be 70% [32]. The numerical value of $\eta_{collecting}$ and $\eta_{mask}$ can be estimated using Monte-Carlo ray tracing in LightTools. For the all-facet illuminator shown in Fig. 12, $\eta_{collecting}$ is estimated to be 89% and $\eta_{mask}$ is estimated to be 81%. Therefore, the transmission of the all-facet illuminator for a 0.55 NA EUV lithography exposure tool is calculated as:

$$Transmission = 89\% \times 70\% \times 70\% \times 81\% = 35.32\%, \tag{28}$$

The performance of the illuminator for EUV lithography exposure tool is measured by the uniformity of the exposure slit at the mask. The uniformity is formulated as:

$$Uniformity = \left(1 - \frac{I_{max} - I_{min}}{I_{max} + I_{min}}\right) \times 100\%, \tag{29}$$

where $I_{max}$ and $I_{min}$ is the maximum and minimum light intensity within the exposure slit in the scanning direction, respectively. The uniformity is simulated by Monte-Carlo ray tracing in

LightTools and calculated by Eq. (29). The light intensity distribution of the exposure slit at the mask under the six illumination pupil shapes is shown in Fig. 16, and their uniformity are listed in Table 3.

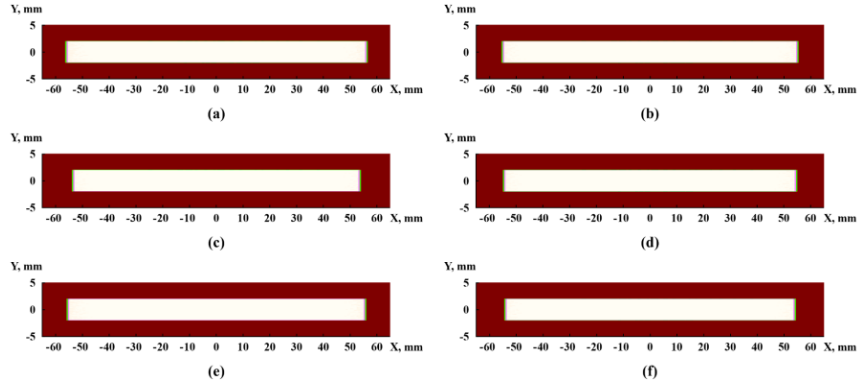

**Fig. 16.** The light intensity distribution of the exposure slit at the mask under the six illumination pupil shapes. (a) Annular illumination pupil shape. (b) Dipole illumination pupil shape. (c) Dipole 90° illumination pupil shape. (d) Quasar illumination pupil shape. (e) Quasar 45° illumination pupil shape. (f) Leaf illumination pupil shape.

Table 3 The uniformity under the six illumination pupil shapes.

| Pupil Shape | Uniformity |
|---|---|
| Annular | 99.12% |
| Dipole | 99.07% |
| Dipole 90° | 99.01% |
| Quasar | 99.04% |
| Quasar 45° | 99.09% |
| Leaf | 99.03% |

The comparisons of the transmission and uniformity between multiple illuminators for EUV lithography exposure tool are listed in Table 4. In conclusion, the illuminator designed using the proposed methods achieves high transmission and high uniformity that are higher than other illuminators.

Table 4 Comparisons between multiple illuminators for EUV lithography exposure tool.

| Method | NA | Character | Structure | Transmission | Uniformity | | | |
|---|---|---|---|---|---|---|---|---|
| | | | | | Annular | Dipole | Quasar | Leaf |
| [25] | 0.3 | Isomorphic | Three-mirror | 25.4% | 95.5% (Not specified pupil shape) | | | |
| [19] | 0.55 | Anamorphic | Four-mirror | 23.6% | 99.02% | 99.03% | 99.04% | 99.01% |
| [16] | 0.5 | Anamorphic | Four-mirror | Not reported | 97.53% | 98.07% | 96.75% | 98.21% |
| Our method | 0.55 | Anamorphic | All-facet | 35.32% | 99.12% | 99.07% | 99.04% | 99.03% |

## 5. Conclusion

In this paper, a design method of the all-facet illuminator for high NA EUV lithography exposure tool and a matching method based on deep RL for the double facets are proposed. The all-facet illuminator is designed using matrix optics and ray tracing, and removing relay system to achieve high transmission. The double facets is matched using the deep RL framework, which includes the policy network consisting of a stochastic module and a deep learning module with improved trainability and low computational demands, and the reward function assigning to three factors affecting uniformity with great optimization direction and fast convergence rate. The double facets matching under a certain illumination pupil shape can be performed automatically, and multiple matching results with high uniformity can be generated rapidly. An all-facet illuminator for a 0.55 NA EUV lithography exposure tool is designed by the proposed method. The simulation results of this EUV illuminator indicate that the transmission is greater than 35%, and uniformity exceed 99% under multiple illumination pupil shapes.

## Funding

## CRediT authorship contribution statement

**Tong Li:** Writing – original draft, Writing – review & editing, Conceptualization, Software, Formal analysis, Visualization. **Yuqing Chen:** Writing – review & editing, Investigation, Data curation. **Yanqiu Li:** Writing – review & editing, Data curation, Supervision, Project administration, Funding acquisition. **Lihui Liu:** Writing – review & editing, Data curation, Supervision.

## Declaration of competing interest

The authors declare that they have no known competing financial interests or personal relationships that could have appeared to influence the work reported in this paper.

## Data Availability

Data will be made available on request.